\documentclass[a4paper,11pt]{article}

\usepackage{tikz} 
\usepackage{jcappub,bm,braket,comment}
\usepackage{physics}
\usepackage{xspace}
\usepackage[T1]{fontenc}
\usepackage{mathrsfs}
\usepackage{enumitem}
\usepackage{hyperref}
\usepackage{caption}
\usepackage{bbm}
\usepackage[normalem]{ulem}

\newcommand{\fref}[1]{figure~\ref{#1}}
\newcommand{\Fref}[1]{Figure~\ref{#1}}
\newcommand{\tref}[1]{table~\ref{#1}}

\newcommand{\trefs}[1]{tables~\ref{#1}}

\newcommand{\sref}[1]{section~\ref{#1}}

\newcommand{\eref}[1]{eq.~(\ref{#1})}

\newcommand{\rref}[1]{ref.~\cite{#1}}

\newcommand{\rrefs}[1]{refs.~\cite{#1}}

\newcommand{\aref}[1]{appendix~\ref{#1}}

\newcommand{\amplSymbol}{\ensuremath{A}}
\newcommand{\orthoAmplSymbol}{\ensuremath{X}}

\newcommand{\ampl}{\ensuremath{\Acal^2 \xi_0}}

\newcommand{\orthoAmpl}{\ensuremath{\xi_0^2 / \Acal}}

\newcommand{\cmb}{{\mathrm{cmb}}}

\newcommand{\deepsphere}{{\sc DeepSphere}\xspace}

\newcommand{\nhat}{\hat{n}}
\newcommand{\svec}{\vec{s}}
\newcommand{\khat}{\hat{k}}

\newcommand{\Acal}{\mathcal{A}}

\newcommand{\alphaem}{\alpha_{\mathrm{em}}}

\newcommand{\Nside}{N_\mathrm{side}}
\newcommand{\Npix}{N_\mathrm{pix}}

\title{\centering 
Extracting Axion String Network Parameters from Simulated \\ CMB Birefringence Maps using Convolutional Neural Networks
}

\author{Ray Hagimoto,}
\author{~Andrew J. Long,}
\author{~and~Mustafa A. Amin}
\affiliation{Department of Physics and Astronomy, Rice University, Houston, TX 77005, USA}

\emailAdd{rmh14@rice.edu}
\emailAdd{andrewjlong@rice.edu}
\emailAdd{mustafa.a.amin@rice.edu}

\abstract{
    Axion-like particles may form a network of cosmic strings in the Universe today that can rotate the plane of polarization of cosmic microwave background (CMB) photons.  Future CMB observations with improved sensitivity might detect this axion-string-induced birefringence effect, thereby revealing an as-yet unseen constituent of the Universe and offering a new probe of particles and forces that are beyond the Standard Model of Elementary Particle Physics.  In this work, we explore how spherical convolutional neural networks (SCNNs) may be used to extract information about the axion string network from simulated birefringence maps.  We construct a pipeline to simulate the anisotropic birefringence that would arise from an axion string network, and we train SCNNs to estimate three parameters related to the cosmic string length, the cosmic string abundance, and the axion-photon coupling.  Our results demonstrate that neural networks are able to extract information from a birefringence map that is inaccessible with two-point statistics alone (i.e., the angular power spectrum).  We also assess the impact of noise on the accuracy of our SCNN estimators, demonstrating that noise at the level anticipated for Stage IV (CMB-S4) measurements would significantly bias parameter estimation for SCNNs trained on noiseless simulated data, and necessitate modeling the noise in the training data. 
}

\begin{document} 
\maketitle
\flushbottom

\section{Introduction}
\label{sec:intro}

Precision measurements of the cosmic microwave background (CMB) radiation yield a wealth of data with which cosmologists are able to infer the constituents of the cosmos~\cite{Planck:2018vyg}. 
The CMB's statistical properties provide compelling evidence for the presence of new physics, beyond the Standard Model of Particle Physics, such as dark matter and dark energy. 
The next generation of CMB telescopes is expected to reach unprecedented levels of precision, particularly in regard to polarization measurements~\cite{CMB-HD:2022bsz}.
These measurements will provide an exciting opportunity to probe signatures of additional beyond the Standard Model physics that are inaccessible with current sensitivity levels~\cite{Chang:2022tzj}.

In this work, we are interested in the cosmological signatures of hypothetical axion-like particles (ALPs) coupled to electromagnetism.  
We assume the standard Chern-Simons interaction, which takes the form 
\begin{align}
    \mathcal{L}_{\mathrm{int}} = -\frac{1}{4} g_{a \gamma \gamma} a F_{\mu \nu} \tilde{F}^{\mu \nu} 
    \;,
\end{align}
where $g_{a\gamma\gamma}$ is the axion-photon coupling parameter, $a(x)$ is the pseudoscalar axion field, $F_{\mu\nu}(x)$ is the electromagnetic field strength tensor, and $\tilde{F}^{\mu\nu}(x)$ is the dual tensor.  
The coupling depends on the fine structure constant $\alphaem \approx 1 / 137$, the axion decay constant $f_a$, and the electromagnetic anomaly coefficient $\Acal$ as $g_{a\gamma\gamma} = \Acal \alphaem / (4 \pi f_a)$.
ALPs arise naturally in string theory as a consequence of the additional compact spatial dimensions~\cite{Svrcek:2006yi,Arvanitaki:2009fg}. 
In these theories, instanton effects lift the ALP potential and their exponential sensitivity leads to a vast range of ALP masses spanning from nearly the Planck scale to far below the current Hubble scale $H_0 \approx 10^{-33}\;\mathrm{eV}$~\cite{Hui:2016ltb,Gendler:2023kjt}. Such ALPs are also expected to interact with electromagnetism at a strength that can possibly be probed with astrophysical and cosmological observations~\cite{Gendler:2023kjt}. 

One of the most well-studied cosmological signatures of ALPs is cosmic birefringence.  
As a linearly-polarized electromagnetic wave propagates through a varying ALP field, the plane of polarization is rotated by an angle $\alpha \propto g_{a\gamma\gamma} \Delta a$ that depends on the Chern-Simons coupling and the change in the ALP field~\cite{Carroll:1991zs,Harari:1992ea,Carroll:1998bd,Lue:1998mq,Liu:2006uh,Liu:2016dcg,Fedderke:2019ajk,Fujita:2020ecn}.  
Measuring cosmic birefringence is an important science driver for current and future CMB experiments~\cite{CMB-S4:2016ple,CMB-HD:2022bsz,Chang:2022tzj}.  
Measurements of isotropic and anisotropic birefringence in upcoming CMB surveys are expected to improve by at least an order of magnitude~\cite{Pogosian:2019jbt,CMB-HD:2022bsz,BICEPKeck:2024cmk}.
An exciting development in recent years is that a measurement of isotropic birefringence in CMB data has been reported with $\approx 3\sigma$ statistical significance~\cite{Minami:2020odp,Diego-Palazuelos:2022dsq,Eskilt:2022wav,Eskilt:2022cff}.
See \rref{Komatsu:2022nvu} a review of recent developments in the measurement of isotropic birefringence.

Various studies have explored the implications of axion-induced birefringence for cosmologically distance sources of polarized light like the CMB.  
While a homogeneous ALP field could induce isotropic cosmic birefringence~\cite{Harari:1992ea,Lue:1998mq,Luo:2023cxo}, there is theoretical motivation to consider configurations such as string and domain wall networks that can form from phase transitions in the early Universe~\cite{Kibble:1976sj}. 
A remarkable feature of birefringence from string loops is that it does not directly depend on string tension and arises even if the string network is a subdominant component of the total energy budget of the universe \cite{Agrawal:2019lkr}. 
Moreover, the anisotropic birefringence signal from a network, if detected, is likely to be more robust against calibration errors that might affect isotropic birefringence measurements. 

Dynamics of topological axion defects and the birefringence induced by them have been studied in \rrefs{Huang:1985tt,Harvey:1988in}. 
CMB birefringence power spectra and non-Gaussian signatures from axion strings have been computed in \rrefs{Agrawal:2019lkr,Jain:2021shf,Hagimoto:2023tqm}.
Constraints on axion-string parameters using published birefringence power spectra have been derived in \rrefs{Yin:2021kmx,Jain:2022jrp}.
The potential of using radio emissions from spiral galaxies to probe birefringence caused by axion strings has been explored in \rref{Yin:2024fez}.
A tomographic constraint on anisotropic birefringence generated at reionization was provided in \rref{Sherwin:2021vgb,Namikawa:2024sax}. 
The birefringence from axion domain walls has been studied in \rrefs{Takahashi:2020tqv,Gonzalez:2022mcx,Kitajima:2022jzz,Ferreira:2023jbu} and axion dark energy in \rref{Obata:2021nql}.  

Recent advances in statistical learning and computing have made it possible to harness the power of neural networks in cosmology~\cite{Schmelzle:2017vwd,Peel:2018aei,Ribli:2019wtw,Fluri:2019qtp,Hortua:2019ryu,Guzman:2021ygf,Taylor:2023deh,Zhong:2024qpf,DES:2024xij}. 
In particular, spherical convolutional neural networks (SCNNs) designed for data with a spherical topology have been developed, such as those implemented in the Python package \deepsphere ~\cite{deepsphere_iclr, deepsphere_rlgm}.
These networks have demonstrated capability in distinguishing cosmological models using simulated weak lensing all sky maps~\cite{deepsphere_cosmo}, and have been used for cosmological parameter inference in KiDS-1000 weak lensing maps~\cite{Fluri:2022rvb}.

\begin{figure}[t]
    \centering
    \includegraphics[width=0.8\linewidth]{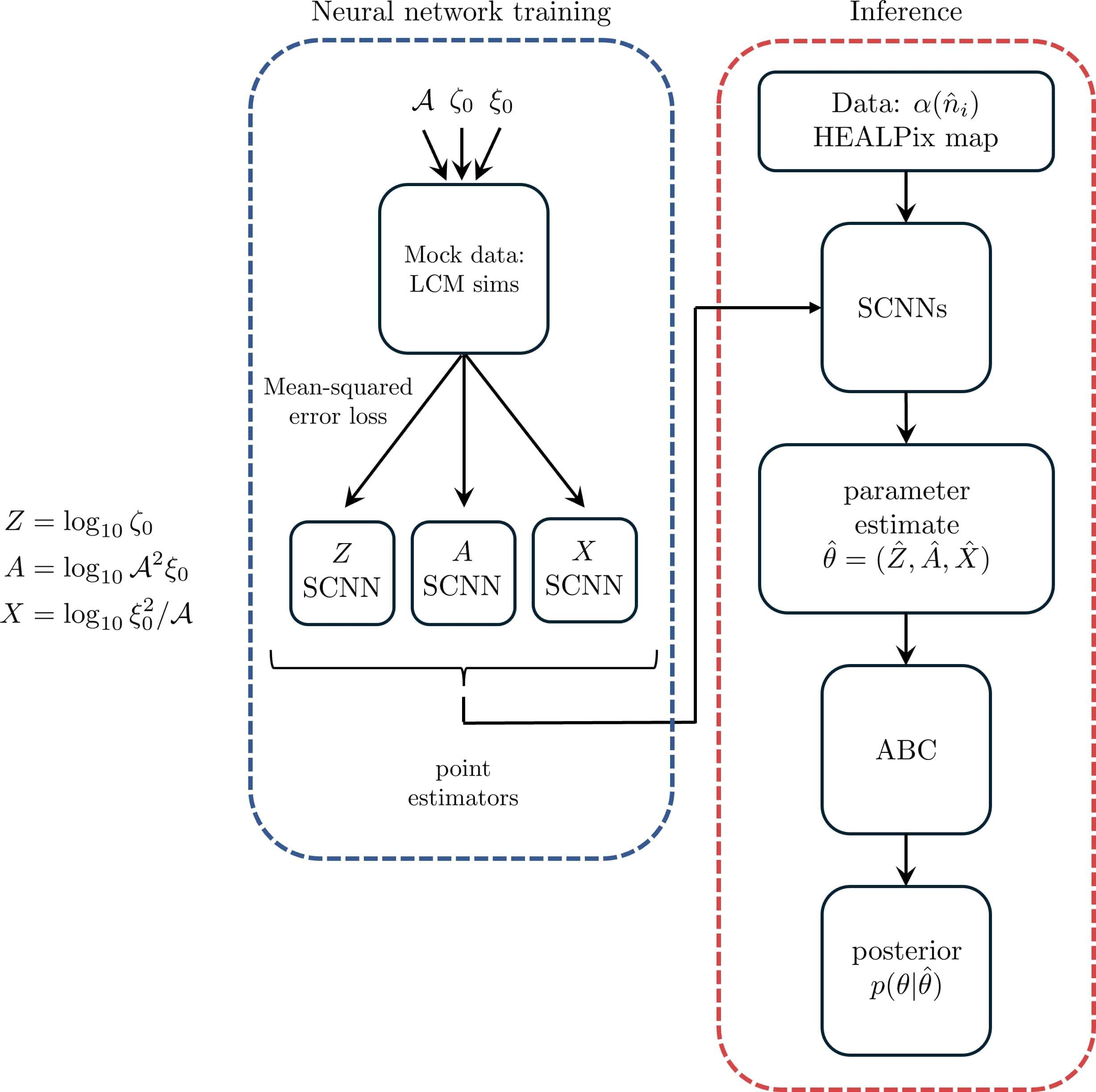}
    \caption{\label{fig:pipeline}
    Illustration of our neutral network training and inference pipelines.  Mock data is generated by performing simulations of the loop crossing model (LCM), which has parameters $\Acal$, $\zeta_0$, and $\xi_0$.  Data takes the form of a pixelated map  of birefringence rotation angles $\alpha(\nhat_i)$ in HEALPix format.  Mock data is used to train three spherical convolutional neural networks (SCNNs), which return point estimates of $Z = \log_{10}(\zeta_0)$, $A = \log_{10} (\ampl)$, and $X = \log_{10} (\orthoAmpl)$.  We validate the training of the SCNNs using two checks of their parameter inference.  One method is approximate Bayesian computation (ABC), which calculates the posterior over the LCM model parameters.  
    }
\end{figure}

In this work we explore how SCNNs can be used to estimate axion string network parameters from simulated all sky maps of CMB birefringence.
To this end, we train SCNNs to estimate the parameters of a phenomenological model known as the loop-crossing model (LCM)~\cite{Jain:2021shf}, given simulated noiseless CMB birefringence maps.
In previous works, measured birefringence power spectra have been compared against the power spectrum predicted from the LCM~\cite{Yin:2021kmx,Jain:2022jrp}, with the tightest constraint yielding $\Acal^2 \xi_0 \lesssim 0.93$ at 95\% confidence level. 
However, a fundamental limitation of this approach is that some model parameters are degenerate at the level of the power spectrum since  in the LCM, the birefringence power spectrum is directly proportional to $\Acal^2 \xi_0$.
The LCM parameter $\xi_0$ is a phenomenological parameter that describes the energy density of the string network in a Hubble volume. 
This means that inference using only the power spectrum is fundamentally unable to independently measure $\Acal$ and $\xi_0$.
To address this issue, the use of higher order statistics such as the trispectrum and wavelet scattering transform have been explored and were shown to break the degeneracy between $\Acal$ and $\xi_0$~\cite{Yin:2023vit,Hagimoto:2023tqm}. 
The ability for SCNNs to learn statistical properties of image data motivates the exploration of this tool for axion string parameter estimation.

Our strategy in this work is to train three spherical convolutional neural networks to estimate the LCM parameters $\zeta_0$, $\Acal$, and $\xi_0$ using realizations of axion-string induced birefringences maps as training data. 
We then assess their performance by using approximate Bayesian computation to sample the posterior distributions. 
This pipeline is illustrated in~\fref{fig:pipeline}.
Our results demonstrate that neural networks are a powerful tool that can be used to look for evidence of cosmic axion strings in future CMB polarization measurements.
In addition to considering noiseless simulations, we explore the effect that adding noise to the maps has on the estimates produced by these neural networks.

\section{Mock birefringence data simulation procedure}
\label{sec:loop-crossing-model}

In this section we discuss how we generated mock data of the anisotropic birefringence arising from an axion string network by employing Loop Crossing Model (LCM) simulations.
The LCM~\cite{Jain:2020dct} treats all axion strings in the network to be circular planar loops whose positions are statistically homogeneous, whose orientations are statistically isotropic, and whose mean abundance and typical size evolve to scale with the cosmological expansion.  
The LCM is informed by numerical 3D lattice simulations of axion string network dynamics~\cite{Yamaguchi:1998gx,Yamaguchi:2002sh,Hiramatsu:2010yu,Hiramatsu:2012gg,Kawasaki:2014sqa,Lopez-Eiguren:2017dmc,Gorghetto:2018myk,Hindmarsh:2019csc,Gorghetto:2020qws,Hindmarsh:2021vih}, which reveal that the network scales with the cosmological expansion (up to a possible logarithmic correction that remains under debate).

\begin{figure}[t]
    \centering
    \includegraphics[width=0.9\linewidth]{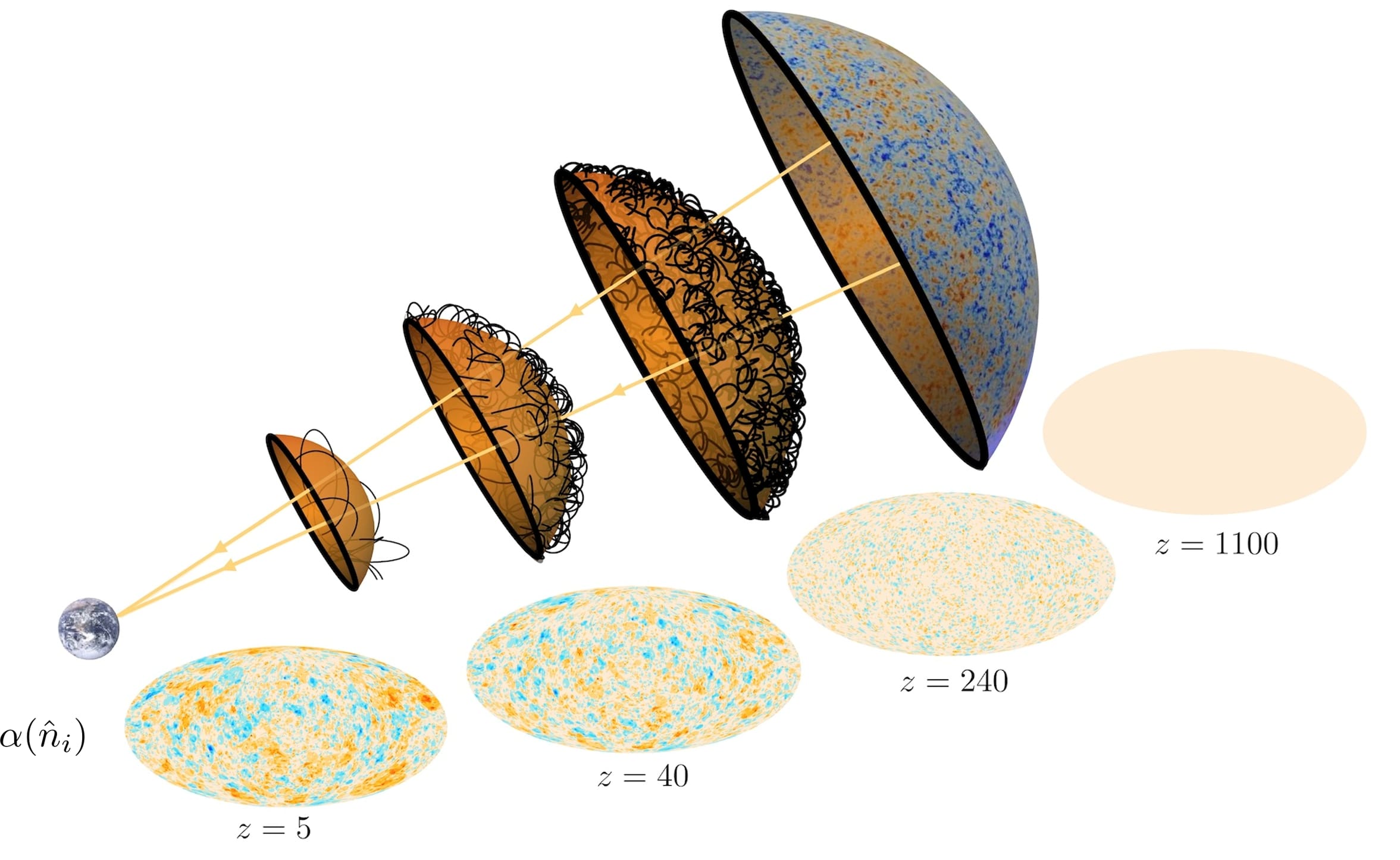}
    \caption{\label{fig:lcm-illustration}
    Illustration of birefringence accumulation in the loop crossing model (LCM). CMB photons (yellow arrowed lines) propagate from the surface of last scattering ($z = 1100$) to Earth ($z = 0$).  The intervening space is filled with an LCM string network consisting of circular planar loops with statistically homogeneous positions and statistically isotropic orientations.  Each time a photon passes through a loop its plane of polarization incurs a rotation of $\Delta \alpha \pm \Acal \alphaem$ depending on the orientation of the loop.  Birefringence accumulates over time with multiple loop crossings.  Three redshift slices ($z=240$, $z=40$, and $z=5$) are illustrated, showing a possible realization of the string network (orange shells with black circles) and the accumulated  birefringence (mollweide projections).  This graphic is figure 4 of \rref{Jain:2022jrp}, and we reproduce it here with permission from the authors. 
    }
\end{figure}

In our implementation, the LCM has four parameters:  a size parameter $\zeta_0$ related to the radius of string loops, an abundance parameter $\xi_0$ related to the number of string loops, the mass parameter $m_a$ related to the string network collapse, and an intensity parameter $\Acal$ related to the amplitude of birefringence.
At time $t$ we assume that all loops in the network have the same (physical) radius $r(t) = \zeta_0 \, d_H(t)$, which grows with time to scale with the increasing Hubble distance $d_H(t) = 1 / H(t)$.  
The average number density of loops also decreases as $n(t) = \xi_0 d_H^{-3}(t) / (2 \pi \zeta_0)$ to maintain scaling.  
A logarithmic deviation from scaling would correspond to a growth in $\xi_0$ by a negligible factor between recombination and today, which we neglect.
The axion mass $m_a$ controls the time when the string network develops domain walls and collapses, through the relation $m_a \sim 3 H(t)$~\cite{Jain:2021shf}.
In this work we restrict ourselves to masses $m_a \lesssim 4 \times 10^{-33}\;\mathrm{eV}$ so that the string networks survive at least until today.
Photons passing through the disk enclosed by a loop develop a birefringence rotation angle of $\Delta\alpha = \pm \Acal \alphaem$ where $\Acal$ is the electromagnetic anomaly coefficient, $\alphaem \approx 1/137$ is the electromagnetic fine structure constant, and the sign depends on the loop's winding number and orientation~\cite{Agrawal:2019lkr}. 
As a photon propagates though multiple loops, its birefringence accumulates $\alpha = \sum \Delta\alpha$.
\Fref{fig:lcm-illustration} is a graphical illustration of the LCM and induced CMB birefringence; we indicate photons propagating through a network of circular planar string loops, shown in three redshift slices.
In the bottom half of \fref{fig:lcm-illustration} we show mollweide projections of the cumulative birefringence map from $z = 1100$ to the indicated redshift.

\begin{table}[t]
    \centering
    \begin{tabular}{c c}
         {\bfseries Parameter}        & {\bfseries Prior}                  \\
         \hline
         $\log_{10}{\zeta_0}$         & $U\big(\log_{10}(0.3), \log_{10}(3)\big)$ \\
         $\log_{10}{\Acal}$         & $U\big(\log_{10}(0.1), \log_{10}(1)\big)$ \\
         $\log_{10}{\xi_0}$         & $U\big(\log_{10}(3), \log_{10}(30)\big)$ \\
    \end{tabular}
    \caption{\label{tab:prior}
    Priors used in inference and mock-data generation. The logarithms are drawn uniformly from the ranges shown. 
    }
\end{table}

The procedure that we employ to create mock data is the following.
\begin{enumerate}
    \item  Draw a set of LCM model parameters $\zeta_0$, $\Acal$, and $\xi_0$ from the prior distributions in~\tref{tab:prior}. We set $m_a = 0$. 
    These priors are informed by theoretical expectations for the values these parameters may take.
    For example $\Acal$ is a sum over the squared electromagnetic charge of particles in the theory, so $\Acal$ is not expected to be much smaller than an $\mathcal{O}(1)$ number.
    This partially motivates us to take our prior to have support in the region $\Acal \in [0.1, 1]$.
    \item  Select a HEALPix resolution parameter.  For all of the work present in this article, we take $\Nside = 128$ corresponding to $N_\mathrm{pix} = 196,608$ pixels.
    This corresponds to an angular scale of approximately $0.5$ degrees.
    For the range of LCM model parameters that we explore, $\Nside = 128$ is adequate to resolve the smallest loops. Since computation time $\propto \Nside^2$, given our resources, we did not perform a complete analysis with $\Nside > 128$.    
    \item Calculate the average number of loops in the network $\langle N_\mathrm{loops} \rangle$ by using numerical methods to evaluate the integral~\cite{Jain:2021shf}
    \begin{align}
        \langle N_\mathrm{loops} \rangle = 2\, \frac{\xi_0}{\zeta_0} \int_{0}^{z_\cmb} \! \dd z \, H^2(z) (1 + z)^{-3} \, s^2(z) 
        \;.
    \end{align}
    Here $z_\cmb = 1100$ is the fiducial redshift at recombination, $s(z) \equiv \int_0^z H^{-1}(z') \, \dd z'$ is the comoving distance from the observer (at $z=0$) to redshift $z$, and $H(z) \equiv H_0 \sqrt{\Omega_r \, (1+z)^4 + \Omega_m \, (1+z)^3 + \Omega_\Lambda}$ is the Hubble parameter at redshift $z$. We assume an $\Lambda$CDM cosmology with $\Omega_r = 9 \times 10^{-5}$, $\Omega_m = 0.3$, and $\Omega_\Lambda = 0.7$. 
    \item Choose the number of loops in this realization by sampling $\hat{N}_\mathrm{loops} \sim \mathrm{Poisson}(\langle N_\mathrm{loops} \rangle)$.
    \item To populate the network with loops, for each loop we generate a random position drawn uniformly from a 2-sphere, a random orientation drawn uniformly from a 2-sphere, a random winding number drawn uniformly from $\pm 1$, and a random redshift $z$ drawn from the probability density (see \aref{app:redshift-derivation})
    \begin{align}
        \label{eq:redshift-pdf}
        p(z) = \frac{H^2(z) (1 + z)^{-3} \, s^2(z)}{\int_{0}^{z_\cmb} \! \dd z' \, H^2(z') (1 + z')^{-3} s^2(z')}
    \end{align}
    with support only on the interval $z \in (0, z_\cmb)$. 
    A derivation of~\eref{eq:redshift-pdf} can be found in~\aref{app:redshift-derivation}.
    Once a random $z$ is chosen, the circular loop's comoving radius is taken to be $r_\mathrm{co} = \zeta_0 (1 + z) / H(z)$.
    \item Using a HEALPix discretization scheme~\cite{Gorski:2004by}, for each loop find the pixels whose line of sight vectors pass through the interior of the loop. At these pixels increment their values by $\Delta \alpha = \pm \Acal \alphaem$.
\end{enumerate}
The result of this procedure is a spatially-discretized birefringence map $\alpha_i = \alpha(\nhat_i)$ that takes values on each of the pixels $\nhat_i$.

\section{Neural network architecture and training}
\label{sec:nn-arch}

Our mock data takes the form of an all-sky birefringence map, having a spherical topology. 
We therefore choose to use a neural network architecture which appropriately accounts for the geometry of the data. 
To this end, we use the \deepsphere Python package,\footnote{
There are several reasons underlying our decision to use \deepsphere: it is suitable for data with spherical topology, it has native support for HEALPix that we use to generate mock data, it has been used in prior cosmological model classification, and its architecture learns directly in position space.  In regard to this last point, since our signal consists of disk-shaped features in position space, we expect that \deepsphere is well suited to the task.   Some alternatives to \deepsphere, such as $S^2\mathrm{CNN}$ \cite{Cohen:2018ldq}, use spherical harmonic transforms.  Since the features can be sharp in position space, the information can be spread over many modes in harmonic space (along with potential numerical errors from the transforms at high multipoles). As a result, we expected that this approach would be less adept. Moreover, $S^2$CNN is expected to be computationally more expensive, see \rref{deepsphere_cosmo}.
} 
which is a library for creating SCNNs~\cite{deepsphere_rlgm,deepsphere_iclr}. 
Previous work has applied \deepsphere to cosmological mock data with HEALPix~\cite{deepsphere_cosmo}; and to KiDS-1000 weak lensing maps for parameter inference~\cite{Fluri:2022rvb}.

We train three neural networks, such that each of them is an estimator for one of the three LCM model parameters.  
Rather than directly learning $\zeta_0$, $\xi_0$, and $\mathcal{A}$, we find that it is advantageous for the networks to instead learn 
\begin{equation}
    \label{eq:ZAX}
    Z \equiv \log_{10}(\zeta_0) 
    \ , \quad 
    A \equiv \log_{10}(\ampl) 
    \ , \quad \mathrm{and} \quad 
    X \equiv \log_{10}(\orthoAmpl) 
    \;.
\end{equation}
This is the case for two reasons. 
First, the birefringence angular power spectrum in the LCM is directly proportional to $\ampl$ and its shape is controlled by $\zeta_0$~\cite{Jain:2020dct}.
This means that power spectrum information can be used to infer these parameters. 
On the other hand, $\orthoAmpl$ is orthogonal to $\ampl$ in the $(\log_{10} \xi_0, \log_{10} \Acal)$ plane so inferring this combination of parameters requires information beyond the power spectrum.
Hence, training a neural network to estimate $\orthoAmpl$ will allow us to assess whether or not information beyond the power spectrum is being recovered.
Second, since $\ampl$ and $\orthoAmpl$ vary over several orders of magnitude, 
and because we care more about the neural networks' ability to provide accurate estimates to within an order of magnitude, training them to learn the base-10 logarithms of these parameters is a direct way to enforce this. 

We use the architectures shown in \trefs{tab:log-zeta-arch},~\ref{tab:log-A2xi0-arch},~and~\ref{tab:log-xi0_A-arch} which can be found in~\aref{app:nn-archs}. 
We use 3 convolutional layers for $\zeta_0$ with a total of 9,881 parameters since we found that this was sufficient for good performance on the range of parameters allowed by our priors (see ~\tref{tab:prior}).
For the $\amplSymbol$ and $\orthoAmplSymbol$ networks
we used deeper networks with 6 to 8 convolutional layers and 20,113 to 3,566,273 trainable parameters respectively.
The convolutions themselves are implemented using \deepsphere's \texttt{ChebyshevConv} layers which approximate the convolution kernels as a Chebyshev polynomial expansion in terms of the discrete Laplacian operator as explained in \rrefs{2013ISPM...30...83S,deepsphere_rlgm}.
These convolutions are approximately equivariant under rotations of the input map, i.e., a rotation of the map followed by a convolution is the same as a convolution followed by a rotation.
After all the convolutions we use global average pooling~\cite{2013arXiv1312.4400L} to ensure that the outputs of the neural networks are invariant under rotations of the input map.

For our mock data set we generate 20,000 axion-string-induced birefringence maps (with HEALPix pixelization) by performing repeated LCM simulations using the parameters $\zeta_0$, $\xi_0$, and $\Acal$ sampled from the priors in~\tref{tab:prior} and by following the procedure described in \sref{sec:loop-crossing-model}.
These data are then split into training and validation sets with a ratio of 80:20.
Finally, we train three neural networks to provide estimators for $Z$, $A$, or $X$ using a mean-squared-error loss~\cite{james2013introduction}. 
Training was performed using the Adam optimizer~\cite{kingma2014adam} with a learning rate schedule starting at $0.005$, which decays by $4\%$ every epoch.   
Training is stopped when the validation loss does not improve by at least $1 \times 10^{-5}$ over $8$ consecutive epochs.
We define the trained neural network as the set of network parameters with the lowest validation loss.

\begin{figure}[t]
    \centering
    \includegraphics[width=0.9\linewidth]{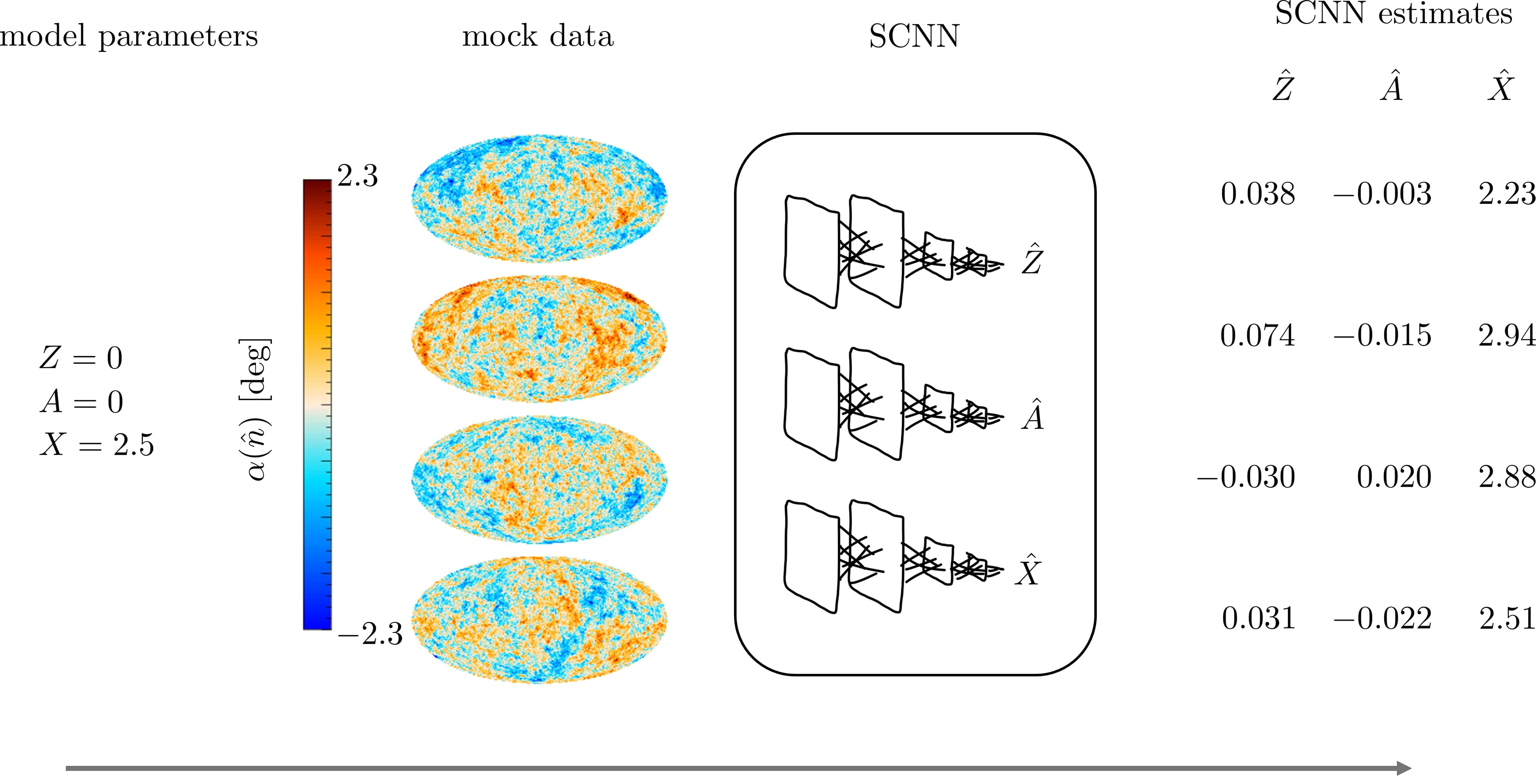}
    \caption{\label{fig:nn-diagram}
    Illustration of LCM parameter estimation using SCNNs on simulated birefringence maps. From left to right we show, (1) a set of LCM model parameters $Z = 0$, $A = 0$, and $X = 2.5$, corresponding to $\zeta_0 = 1$, $\Acal = 0.316$, and $\xi_0 = 10$; (2) four realizations of birefringence mock data generated using LCM simulation; (3) graphical depiction of our three SCNNs; and (4) parameter estimates furnished by each of the three SCNNs for each of the four maps. Note that the graphical depiction does not faithfully illustrate the architecture of an SCNN; \deepsphere does not represent the convolutional kernels as 2D arrays.
    }
\end{figure}

To illustrate the performance of the three networks, we show an example in \fref{fig:nn-diagram}.  
For this example, we take the LCM model parameters to be $\zeta_0 = 1$, $\Acal = 0.316$, and $\xi_0 = 10$, which correspond to $Z = 0$, $A \approx 0$, and $X \approx 2.5$.
We generate four pixelated birefringence maps $\alpha(\nhat_i)$, which are random realizations of the LCM simulation procedure.  
These maps are then passed to the three SCNNs, which have already been trained, yielding estimates $\hat{Z}$, $\hat{A}$, and $\hat{X}$.
In general we will use hatted variables to denote the outputs of the SCNNs, and unhatted variables to denote the LCM model parameters.
We emphasize that the neural network is deterministic, such that the same map $\alpha(\nhat_i)$ always yields the same estimate, $\hat{Z}$ for example.
However, the procedure of generating the mock data via the LCM simulation is stochastic, so a single set of LCM model parameters generates many possible realizations $\alpha(\nhat_i)$.
In this example, one can see that two of the SCNNs are performing very well; the $\hat{Z}$ and $\hat{A}$ estimates are close to the input model parameters $Z$ and $A$.  
The third SCNN performs moderately well; the estimate $\hat{X}$ differs from the input parameter $X = \log_{10}(\orthoAmpl) = 2.5$ by as much as $\Delta X = 0.44$, which corresponds to a factor of $2.8$ in $\orthoAmpl$.  
We discuss the SCNN performance further in the next section, where we also provide quantitative measures of success.  

\section{Validation of neural network performance}
\label{sec:inference}

In order to assess the performance of the neural networks to provide accurate and precise estimates of the LCM model parameters, we perform the following two tests.
To quantify the networks' accuracy, we directly compare SCNN parameter estimators with true LCM model parameters across a 2D slice of the parameter space.  
To quantify the networks' precision, we employ approximate Bayesian computation to sample the posterior distribution and infer the spread in the network output.  
In the following subsections, we discuss both approaches.  

\subsection{Estimator performance on known inputs}
\label{sec:results-heatmap}

To get a sense of the reliability of the parameter estimates from our neural networks we want to quantify deviations from known inputs.
This can be achieved by scanning the parameter space and calculating the average error at each point.
To do this calculation we perform 992,600 draws of the LCM model parameters $\zeta_0$, $\xi_0$, and $\Acal$ from the priors in~\tref{tab:prior}, and calculate the corresponding $Z$, $A$, and $X$ using \eref{eq:ZAX}.
For each draw we do an LCM simulation and pass the mock data to the neural networks which return parameter estimates $\hat{Z}$, $\hat{A}$, and $\hat{X}$. 
We then divide the $(A, X)$ plane into $50 \times 50 = 2500$ bins.
In each bin we compute the error magnitude defined as 
\begin{align}\label{eq:error_magnitude}
    \text{error magnitude} = \text{average of }  \sqrt{(\hat{A} - A)^2 + (\hat{X} - X)^2}
    \;,
\end{align}
where the average is with respect to the samples in the bin.
Some bins have no samples, because they are excluded by our priors; other bins contain between 65 and 809 samples.

\begin{figure}[t]
    \centering
    \includegraphics[width=0.85\linewidth]{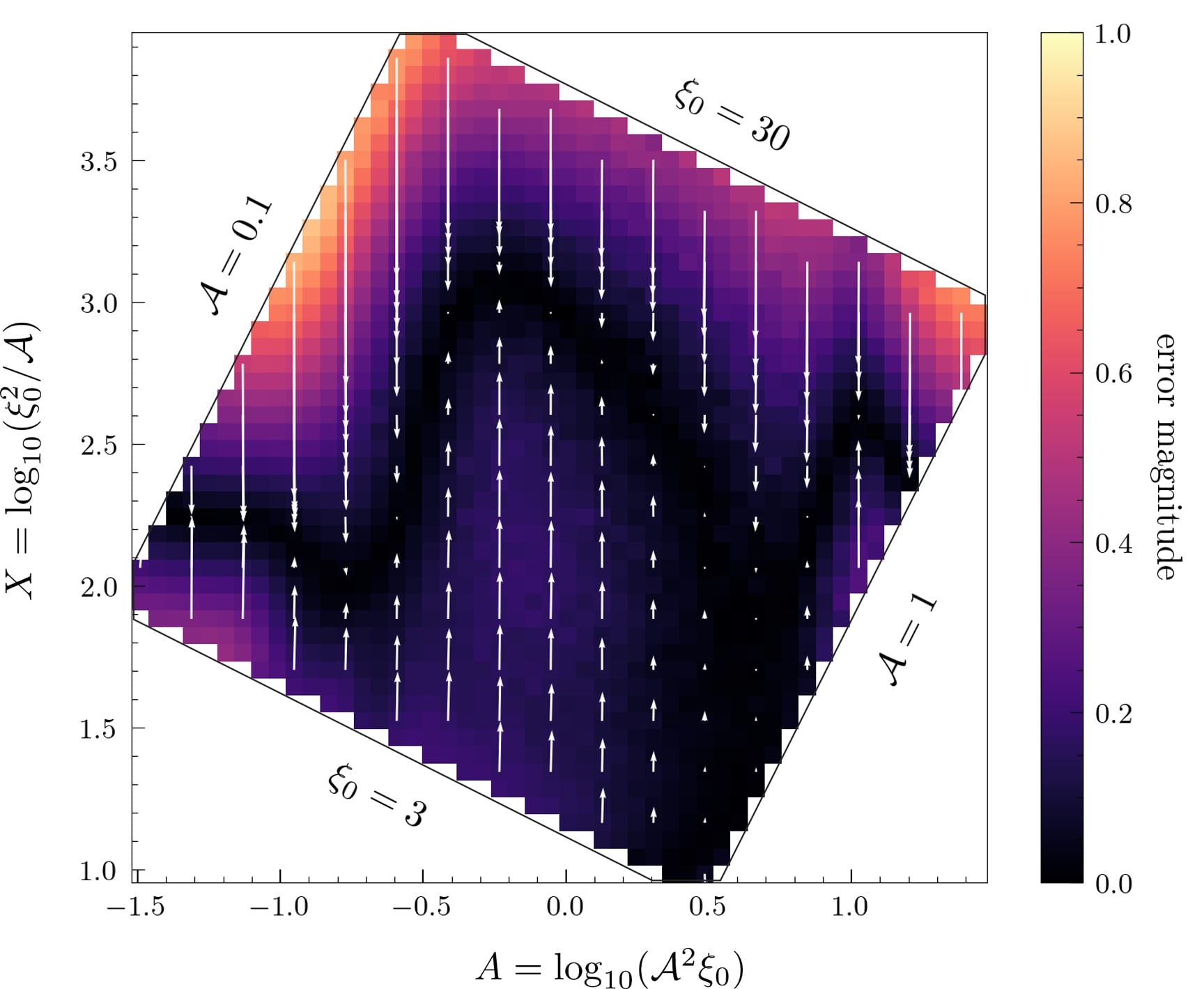}
    \caption{\label{fig:heatmap}
    An illustration of the performance of our trained SCNNs.  We show the error magnitude \eqref{eq:error_magnitude} as a colored heatmap where cooler/darker colors indicate better performance.  We also show the displacement from input parameter pair to the average output of the SCNNs as a white arrow.  There are no samples in the white regions, which are outside of our prior range, as indicated by the diagonal labels; see also \tref{tab:prior}.
    }
\end{figure}

The results of this analysis are shown in \fref{fig:heatmap}.
The color heat map shows the error magnitude in each bin. 
The white regions have no samples, because they are outside of our prior range, as indicated by the labeled values of $\Acal$ and $\xi_0$. 
The arrows indicate the displacement from the true value to the average estimated value. 
For example, an arrow pointing upward means that the neural networks tend to overestimate the value of $X$ but are relatively accurate for $A$. 

This figure offers several indications of the performance of the SCNNs.
The error magnitude is typically smaller than $0.4$ across most of the $(A,X)$ parameter space, corresponding to a factor of $10^{0.4} \approx 2.5$ in the parameters $\ampl$ and $\orthoAmpl$.  
However, the error predominantly arises from $\orthoAmpl$, as indicated by the mostly-vertical arrows.
This poorer performance in the $X = \log_{10}($\orthoAmpl$)$ estimator was expected, and we discuss it further in \sref{sub:interpretation}.

Notice that the region of parameter space in which $X$ is large tends to have larger error magnitude (i.e., warmer colors).  
For larger $\xi_0$ the error is expected to grow, because the network contains a greater number of loops.  
The birefringence map generated by many overlapping loops appears increasingly like a Gaussian random field~\cite{Hagimoto:2023tqm}, and obscures information about the string network. 
This leads to a larger error magnitude. 

The performance of the network to accurately estimate $X$ is particularly poor at the upper-left boundary.  
This is indicated by the bright orange cells on the heat map and the long downward arrows.  
Here the error magnitude reaches $0.8$ corresponding to underestimating $\orthoAmpl$ by a factor of $10^{0.8} \approx 6.3$.

The $X$ neural network tends to overpredict when the input value is small (i.e., upward arrows near the bottom of the figure) and underpredict when the input value is large (i.e., downward arrows near the top).
Consequently, there is a region of parameter space where the predictions are particularly accurate -- this is the very dark region of the plot. 
If the network were trained again using a new set of randomly-generated training data, we anticipate that the general trends seen in \fref{fig:heatmap} would persist, while the particular values of the error magnitude would change.

\subsection{Approximate Bayesian computation}
\label{sec:results-posterior}

We designed the SCNNs to provide a single estimator of the LCM model parameters (rather than a probability density).   
In this section, we discuss how approximate Bayesian computation (ABC)~\cite{fan2018abc} can be employed to ascribe an uncertainty to those estimates.  
More formally, the uncertainty of the network is quantified by the posterior distribution $p(\theta | \hat{\theta})$, where $\theta \equiv (Z, A, X)$ are the parameters and $\hat{\theta} \equiv (\hat{Z}, \hat{A}, \hat{X})$ are the estimates of the parameters. 
To obtain samples from $p(\theta | \hat{\theta})$ we use ABC, following the procedure outlined below.
\begin{enumerate}
    \item Define a distance measure between parameter estimates $\hat{\theta}$ and a target value $\hat{\theta}_\mathrm{target}$ as 
    \begin{align*}
        \rho(\hat{\theta}, \hat{\theta}_\mathrm{target}) 
        = \sqrt{(\hat{Z} - {\hat{Z}_{\mathrm{target}}})^2
        + (\hat{A} - \hat{A}_\mathrm{target})^2
        + (\hat{X} - \hat{X}_\mathrm{target})^2 } 
        \;.
    \end{align*}
    \item Pick a tolerance $\epsilon$.  We use $\epsilon = 0.07$, which was found to be sufficient for convergence.   
    \item Sample LCM model parameters $\zeta_0$, $\Acal$, and $\xi_0$ from the priors in~\tref{tab:prior}, and compute the corresponding $\theta = (Z, A, X)$ using \eref{eq:ZAX}.
    \item Generate a pixelated LCM birefringence map $\alpha(\nhat_i)$ following the procedure in \sref{sec:loop-crossing-model}.  
    \item Pass the map through each neural network to obtain estimates $\hat{\theta} = (\hat{Z}$, $\hat{A}$, $\hat{X})$.
    \item Keep the sampled $\theta$ if $\rho(\hat{\theta}, \hat{\theta}_\mathrm{target}) < \epsilon$.
    \item Repeat steps $3-6$ until a desired number of accepted samples is reached.
\end{enumerate} 

\begin{figure}[t]
    \centering
    \includegraphics{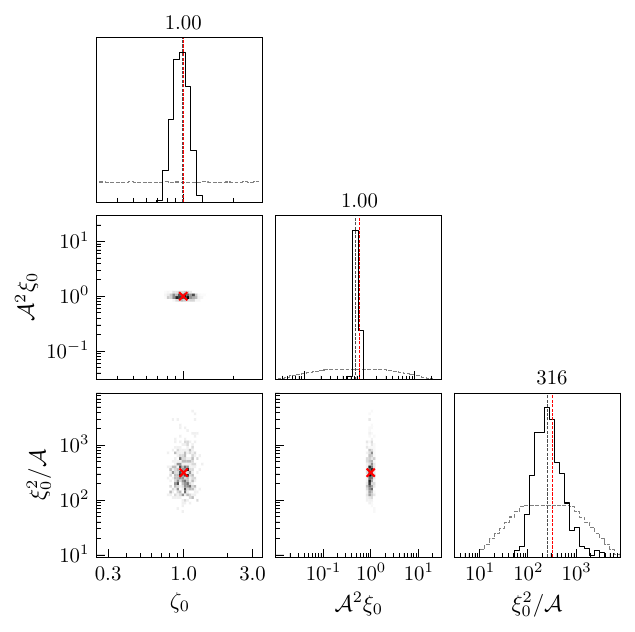}
    \caption{\label{fig:posterior}
    Posteriors on the parameters $Z = \log_{10}(\zeta_0)$, $A = \log_{10}(\ampl)$, and $X = \log_{10}(\orthoAmpl)$ obtained using ABC sampling when $\hat{\theta}_\mathrm{target} = (\hat{Z}$, $\hat{A}$, $\hat{X}) = (0,\,0,\,2.5)$.  The diagonal subplots show the 1D marginal posteriors.  The gray dashed curves in the diagonal subplots depict the prior for each parameter.  Gray vertical lines depict the parameter with the highest posterior density.  Red vertical lines depict $\hat{\theta}_\mathrm{target}$.  Lower-left subplots show a histogram of 2D marginal posteriors.  Red crosses depict the projection of $\hat{\theta}_\mathrm{target}$ in each plane.
    For another example with a different $\hat{\theta}_\mathrm{target}$ see~\aref{app:additional}
    }
\end{figure}

\Fref{fig:posterior} shows the outcome of this procedure for a representative parameter point.  
We show samples from the posterior distribution for $\hat{\theta}_\mathrm{target} = (0, 0, 2.5)$, which corresponds to $\hat{\zeta}_0 = 1$, $\hat{\xi}_0 = 10$, and $\hat{\Acal} = 0.316$. 
The three lower subplots show the 2D marginal posteriors for each pair of parameters, and the three upper subplots show the 1D marginal posterior (black histogram) and prior (gray dashed histogram) of each individual parameter. 
The red cross indicates the target value $\hat{\theta}_\mathrm{target}$.

From the 2D posteriors, we can assess possible correlations between estimator errors. 
No significant correlations are observed. 
Since the networks are trained independently we expect uncorrelated errors. 

From the 1D posteriors, we can assess the SCNNs' uncertainties.  
For each of the three estimators, the posteriors are approximately centered at the target value, which is an indication of the SCNNs' accuracy. 
The posteriors have standard deviations of $\sigma_{Z} = 0.03$, $\sigma_{A} = 0.014$, and $\sigma_{X} = 0.3$. 
This means that the $\zeta_0$ parameter is typically within a factor of $10^{0.03} \approx 1.07$ of the target value; the $\ampl$ parameter is typically within a factor of $10^{0.014} \approx 1.03$ of the target value; and the $\orthoAmpl$ parameter is typically within a factor of $10^{0.3} \approx 2$ of the target value. 
For $\zeta_0$ and $\ampl$ the SCNNs are quite precise with uncertainties below $10\%$ (for this example). 
For $\orthoAmpl$ the uncertainty is much larger.  
Nevertheless, all three 1D posteriors, even $\orthoAmpl$, have standard deviations that are smaller than the prior distribution's as evidenced by the black histograms being narrower than the grey lines in~\fref{fig:posterior}. 

\subsection{Interpretation of results}
\label{sub:interpretation}

The results shown in \fref{fig:heatmap} and \fref{fig:posterior} indicate that the neural networks have learned ways to extract information to infer these parameters. 
This is especially interesting for $X = \log_{10}(\orthoAmpl)$ since part of our motivation for this work was to see if SCNNs can break the parameter degeneracy between $\Acal$ and $\xi_0$ that is present in power spectrum-only analyses~\cite{Yin:2023vit,Hagimoto:2023tqm}. 
The stronger predictive power of the $Z$ and $A$ neural networks over the $X$ network can be understood in the following way: unlike for $Z$ and $A$, there is no information contained in the power spectrum which allow one to discern between different values of $X$. 
Therefore, the only way for the $X$ estimator to learn useful information is to extract information beyond the power spectrum, which is harder because it requires the network to learn more complex patterns. 

\section{Estimator degradation due to noise}
\label{sec:noise}

We trained our neural networks to provide parameter estimates on noiseless maps. 
However, real birefringence maps are reconstructed from CMB polarization data. 
This means that birefringence maps will have noise sourced from the $Q$ and $U$ measurements as well as the statistical estimators used to reconstruct the birefringence.
A popular technique for birefringence reconstruction is to use quadratic estimators~\cite{Kamionkowski:2008fp,Gluscevic:2009mm,Yadav:2009eb}.
For example, POLARBEAR, ACT, and SPT have all used quadratic estimators in their analysis of anisotropic birefringence~\cite{POLARBEAR:2015ktq,Namikawa:2020ffr,SPT:2020cxx}. 
For a more general discussion of CMB birefringence quadratic estimators and their statistical reconstruction noise see~\cite{Yin:2021kmx}.
In this section we explore how our neural networks, which were trained on noiseless maps, can perform on noisy data. 

\begin{figure}[t]
    \centering
    \includegraphics[width=0.95\linewidth]{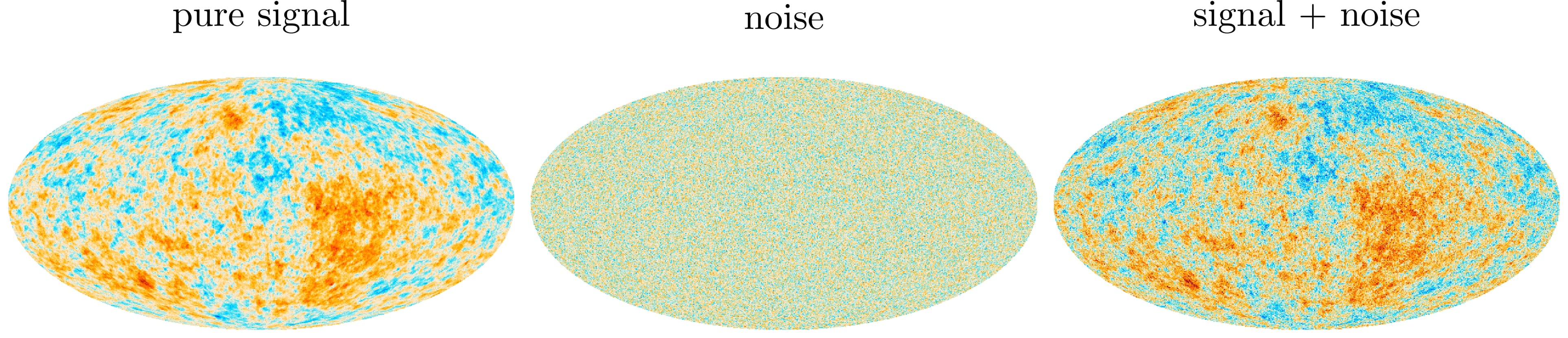}
    \caption{\label{fig:noise-example}
    \emph{Left:} a realization of a CMB birefringence map generated with LCM parameters $\zeta_0 = 1.0$, $\Acal = 0.316$, and $\xi_0 = 10.0$.  \emph{Middle:} a realization of Gaussian random noise for a CMB-S4-like experiment.  \emph{Right:} combined signal and noise.  Since the noise power spectrum follows a white noise profile, the noise level increases toward smaller angular scale.  As a result, the granularity of the noise in the map shown is determined by the resolution of the map.  A higher resolution would give the appearance of smaller scale fluctuations.
    }
\end{figure}

We model the noisy pixelated birefringence maps as $\alpha_S(\nhat_i) + \alpha_N(\nhat_i)$ where the signal $\alpha_S(\nhat_i)$ is generated from the LCM simulation.  
The noise $\alpha_N(\nhat_i)$ is assumed to be a Gaussian random field drawn from the angular power spectrum $N_\ell$, uncorrelated with the signal.  
We assume $N_\ell \propto \ell^0$ is constant, corresponding to white noise.  
This choice is motivated by the noise being dominated by reconstruction noise~\cite{Yin:2021kmx}. 
We study different noise levels by multiplying the expected CMB-S4 noise level $N_\ell^\mathrm{CMB-S4} = 1.5 \times 10^{-5}\ \mathrm{deg}^2$, which we obtained from figure 1 of~\rref{Yin:2021kmx}, by a multiplicative factor.  
We use the \texttt{synfast} method of \texttt{healpy} to generate $\alpha_N(\nhat_i)$.  
An example of the signal and noise maps appears in \fref{fig:noise-example}.  

\begin{figure}[t]
    \centering
    \includegraphics[width=0.95\linewidth]{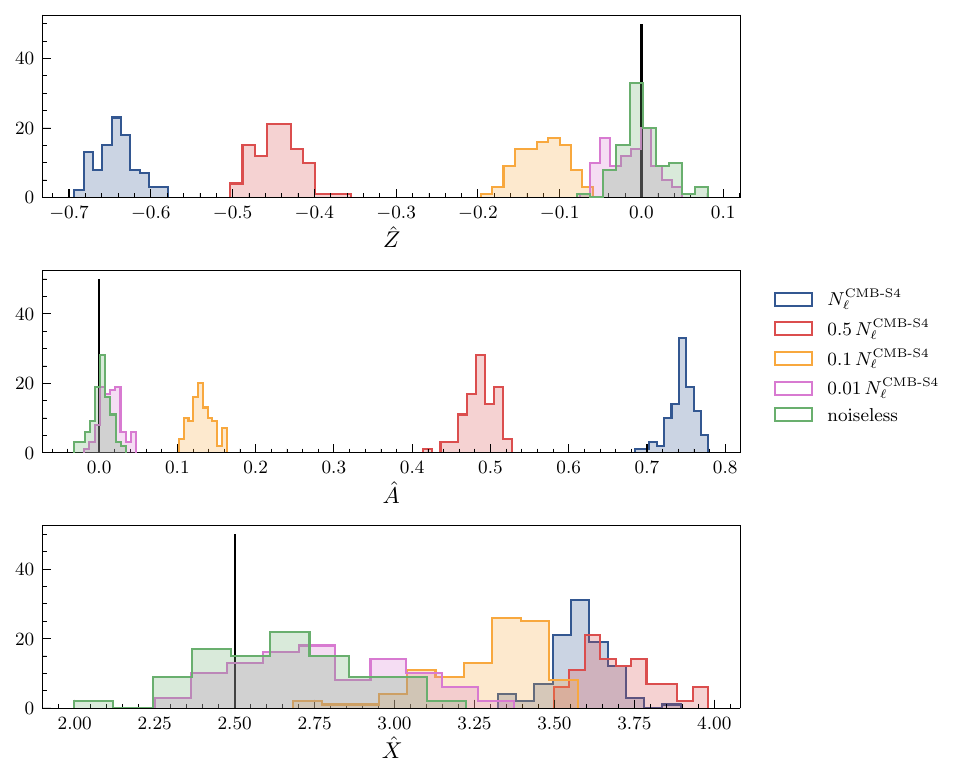}
    \caption{\label{fig:noise}
    Performance of our SCNNs on noisy mock data.  For each noise level (colored histograms) we show the sample distribution over the SCNN estimators $\hat{Z}$ (top), $\hat{A}$ (middle), and $\hat{X}$ (bottom).  To generate each histogram, we perform 100 LCM simulations with $\zeta_0 = 1$, $\xi_0 = 10$, and $\Acal = 0.316$ (corresponding to $Z = 0$, $A = 0$, and $X = 2.5$, indicated by the vertical black bar), add white noise up to the level of a CMB-S4-like experiment, and pass these noisy birefringence maps to our SCNNs, which were trained on noiseless mock data.  
    }
\end{figure}

The histograms in \fref{fig:noise} show the distribution of estimators for each neural network given LCM parameters $\zeta_0 = 1$, $\Acal = 0.316$, and $\xi_0 = 10$, which correspond to $Z = 0$, $A = 0$, and $X = 2.5$.
The networks were trained on noiseless maps and tested on birefringence maps with various noise levels.
The results for each estimator are shown across three panels.
In each panel we depict the input LCM parameter value with a vertical black line.
In general the estimators tend to give biased predictions with comparable variances when provided with noisy data. 

In order from top to bottom the panels show our results for the $Z$, $A$, and  $X$ estimators.
The $Z$ estimator exhibits a bias toward smaller values than the true parameter at $Z = 0$. 
This is shown by the fact that the centers of the distributions move to smaller values with increasing noise. 
For example, the prediction for $0.5\,N_\ell^\mathrm{CMB-S4}$ is centered at $Z \approx -0.45$, whereas the prediction for $N_\ell^\mathrm{CMB-S4}$ is centered at around $Z \approx -0.65$.
This can be understood as the neural network interpreting the noise as part of the signal.
Since the noise power spectrum is approximately a power law with a positive index over the range of multipoles that can be probed with the map resolution of $\Nside = 128$, the noise looks like many small loops about the size of a pixel.
In a similar way, the $A$ estimator biases its predictions to larger values than the true value.
Again, this is to be expected since the neural network interprets the noise as part of the signal. 
At the level of the power spectrum the noise is additive: $C_\ell^\mathrm{tot} = C_\ell^\mathrm{signal}  + C_\ell^\mathrm{noise}$, which can partially account for the trend in the prediction bias.
The $X$ estimator is the most affected by the addition of noise.
Like the $A$ estimator, its predictions are biased above the true value. 
However, we see that while the bias for the $Z$ and $A$ estimators  is less than an order of magnitude for all noise levels, the $X$ estimator's bias is larger than the other estimators' bias at every noise level. 
This upward bias is expected since $X$ is proportional to $\xi_0^2$, which controls the number density of loops. 
If the noise is interpreted as signal it will have the appearance of adding many loops on small scales, leading to an upward bias.

In order for all three estimators to be biased by no more than an order of magnitude, one would need an experiment with about 10 times less noise than CMB-S4 as shown by the yellow histograms in~\fref{fig:noise}.
This analysis reveals that our noiseless estimators prove inadequate to provide reliable estimates at the noise level of a CMB-S4-like experiment. 
Therefore methods for improving the estimators are required. 
A discussion of some of these can be found in~\sref{sec:summary}.

\section{Summary and conclusion}
\label{sec:summary}
In this work, we have explored how spherical convolutional neural networks (SCNNs) can be used to perform parameter inference on simulated axion-string-induced birefringence maps.
Below is a summary of the main contributions and results of this work:

\begin{itemize}
    \item We developed a pipeline for generating HEALPix maps of anisotropic birefringence based on a simulation of the loop-crossing model (LCM), which has phenomenological parameters $\zeta_0$, $\xi_0$, and $\mathcal{A}$, and  which control the radius of string loops, the number of string loops, and the birefringence accumulated by a photon after passing through a single loop, respectively.
    The LCM parameterization is informed by string network dynamics.  
    
    \item We trained three independent SCNNs to estimate the parameters $Z \equiv \log_{10} \zeta_0$, $A \equiv \log_{10} \ampl$, and $X \equiv \log_{10} \orthoAmpl$ from these maps. 
    The choice of these parameters is driven by the following considerations.
    First, the birefringence power spectrum from axion strings is proportional to $\ampl$, so it is natural to train our SCNNs to learn this combination of parameters and another, $\orthoAmpl$ for which the power spectrum provides no information.
    Second, some parameters can take values over several orders of magnitude. 
    This motivates us to train the SCNNs to infer the base-10 log of the parameters.

    \item We evaluated the SCNN estimators' performance on birefringence maps generated from known LCM model parameters (see \fref{fig:heatmap}), finding that the networks performed very well for $Z$ and $A$, and moderately well for $X$.  
    For example, the $Z$ and $A$ estimators are typically biased by no more than $0.04$ units, which corresponds to a factor of $10^{0.04} \approx 1.10$ deviation from the true value of $\zeta_0$ or $\ampl$. On the other hand, depending on the input value the $X$ estimator is typically  biased by less than $0.3$ units corresponding to a factor of $10^{0.3} \approx 2$ error in $\orthoAmpl$ -- although the error can be as high as a factor of 6 for some values of $\orthoAmpl$.
    
    \item We used approximate Bayesian computation to sample the posterior distribution for the case when the estimators yield $\hat{Z} = 0$, $\hat{A} = 0$, and $\hat{X} = 2.5$, which allowed us to quantify the statistical uncertainty of the predictions (see~\fref{fig:posterior}).
    The estimates for $Z$ and $A$ had standard deviations of $\sigma_{Z} = 0.03$ and $\sigma_{A} = 0.014$, respectively.
    The $X$ estimator had a larger uncertainty with $\sigma_{X} = 0.3$.

    \item By simulating CMB birefringence maps with various noise levels we demonstrated that the accuracy of the neural networks is degraded in a qualitatively predictable way. 
\end{itemize}

Our work was motivated in part by the question: can a neural network learn information about a birefringence map that is inaccessible with its power spectrum alone?  
In particular, in the context of the loop crossing model, the birefringence angular power spectrum only depends upon the intensity and abundance parameters, $\Acal$ and $\xi_0$, through the combination $\ampl$, but not the combination $\orthoAmpl$.  
Based on several tests of their accuracy and precision, we conclude that the SCNNs that learned $\ampl$ and $\orthoAmpl$ are able to furnish reliable estimators of these parameters when provided with noiseless birefringence maps generated from LCM simulations.  
However, the uncertainty in $\orthoAmpl$ is much larger than in $\ampl$, approximately a factor of $2$ versus $3\%$ for the sample parameter point illustrated in \fref{fig:posterior}.  
This was expected, since information about $\orthoAmpl$ is not encoded in the power spectrum, but rather stored in higher-point correlations.  
Nevertheless, our priors based on UV considerations and string network simulations allow the parameters $\Acal$ and $\xi_0$ to each vary by about an order of magnitude, and even the factor of $2$ uncertainty in $\orthoAmpl$ is informative.

Our work demonstrates that even with an unoptimized architecture and inference pipeline, SCNNs are capable of extracting information beyond the power spectrum. 
Our work demonstrates that even with an unoptimized architecture and inference pipeline, SCNNs are capable of extracting information beyond the power spectrum. 
We anticipate that better performance (particularly in extracting $\orthoAmpl$) could be achieved with an alternative architecture and/or extended inference pipeline.  
For example, instead of training a neural network directly on birefringence maps, as done in this work, it might be beneficial to first compress the maps into deep summaries and then construct minimally biased estimators from them. 
For details on creating these estimators and training neural networks to learn deep summaries, see~\rref{Fluri:2021qxx}.

These results are a step toward a pipeline that can be used to do inference on real data, but there are areas available for improvement.
In~\sref{sec:noise} we found that noise introduces additional biases in the predictions, particularly for $X$, which exhibited biases as large as two orders of magnitude for CMB-S4 noise levels.
We found that to yield parameter estimates within an order of magnitude from the true value, one would need an experiment with 10\% of the noise level compared to CMB-S4.
To address noise more effectively, one could explore two potential strategies. 
First, neural networks could (and should!) be trained on maps that already include noise specific to the experimental setup. 
This would allow the model to learn both the signal and the noise features, leading to more robust predictions.
Additionally, one could implement techniques from~\cite{Zhong:2024qpf}, which demonstrated improvements by using a combination of max/average pooling and random permutations in the deeper layers of the network.
These regularization and data augmentation techniques help the network generalize better in noisy environments by preventing overfitting to large-scale correlations that might include noise.
We leave the exploration of these techniques to future work.

\begin{acknowledgments}
This material is based upon work supported (in part: R.H. and A.J.L.) by the National Science Foundation under Grant Nos.~PHY-2114024 and PHY-2412797.  
M.A.A. is supported by a DOE grant DOE-SC0021619.
Some of the results in this paper have been derived using the healpy and HEALPix packages. 
A.J.L. thanks Bhuvnesh Jain and Matthew Johnson for illuminating discussions of machine learning methods.  R.H. would like to thank Siyang Ling for his invaluable contributions to the LCM simulation code, and Juehang Qin, Dorian Amaral, and Ivy Li for useful conversations about statistical learning.

\end{acknowledgments}

\appendix

\section{Loop redshift probability density}
\label{app:redshift-derivation}

This appendix provides a derivation of \eref{eq:redshift-pdf}, which gives the probability density $p(z)$ to find a string loop at redshift $z$.
In the LCM the average number of string loops with comoving radius between $r$ and $r + \dd r$, with comoving position between $\svec$ and $\svec + \dd \svec$ on the observer's past light cone, and with orientation (normal to the plane of the loop) between $\khat$ and $\khat + \dd \khat$ is given by~\cite{Jain:2021shf}
\[
    \dd N = \nu(r, z) \, \dd r \, \dd^3 \svec  \, \frac{\dd^2 \khat}{4 \pi} 
    \;.
\]
The assumption that loops are oriented isotropically implies that $\nu$ is independent of $\khat$, and the assumption that the loops are distributed homogeneously throughout space at a given time implies that $\nu$ only depends on $\svec$ through $s = |\svec|$, which is a proxy for time (or redshift) on the past light cone.
In spherical polar form we have $\dd^3 \svec = s^2 \, \dd^2 \nhat \, \dd s$, where $\nhat$ is the unit vector pointing in the direction of $\svec$.
To write $\dd^3 \svec$ in terms of $z$ note that $s(z) = \int_0^z H^{-1}(z') \dd z'$. 
Hence, $\dd^3 \svec = s^2(z) \, H^{-1}(z) \, \dd z \, \dd^2 \nhat$.

In order to more conveniently parameterize the scaling property of string loop networks we introduce a kernel function $\chi(\zeta, z)$ as in \rref{Jain:2021shf} such that, 
\begin{align}
    \nu(r, z) 
    & = \int_0^\infty \! \dd \zeta \, \chi(\zeta, z) \ \frac{H(z)^2 (1 + z)^{-2}}{2 \pi r} \, \delta\bigl[ r - \zeta (1 + z) / H(z) \bigr] 
    \;.
\end{align}
To implement the assumption that all string loops have the same radius at a given time, we take $\chi(\zeta, z) = \xi_0 \, \delta(\zeta - \zeta_0)$.
Integrating over the loop radius and orientation gives the average number of loops with comoving position between $\svec$ and $\svec + \dd\svec$ to be
\begin{align}
    \dd N 
    \ = \ 
    \int_0^\infty \! \dd r \int_{4\pi} \! \! \frac{\dd^2\khat}{4\pi} \, \nu(r,z) \, \dd^3\svec 
    \ = \ 
    \frac{\xi_0}{2 \pi \zeta_0} \Bigl( H(z)^2 (1 + z)^{-3} \, s^2(z) \, \dd z \Bigr) \dd^2 \nhat \; .
\end{align}
The expression on the right implicitly defines a probability density over $z$ which is obtained by normalizing it over the range of possible redshift values. 
Hence, taking $z_\mathrm{cmb} = 1100$, we have
\begin{align*}
    p(z) =  \frac{H^2(z) (1 + z)^{-3} \, s^2(z)}{\int_0^{z_\mathrm{cmb}}H^2(z) (1 + z)^{-3} \, s^2(z) \, \dd z} 
    \;,
\end{align*}
which appears in \eref{eq:redshift-pdf}. 

\section{Neural network architectures}
\label{app:nn-archs}

This appendix contains the architectures used for our SCNNs in \trefs{tab:log-zeta-arch}-\ref{tab:log-xi0_A-arch}.
To construct our neural networks we use the Python package \deepsphere~\cite{deepsphere_iclr, deepsphere_rlgm}.
This package provides implementations of layers designed for use on HEALPix formatted maps.
These include the \texttt{ChebyshevConv} and \texttt{MaxPool} layers which perform convolutions and pooling. 

All of the computations for this project were performed using an NVIDIA RTX 4070 (Laptop) GPU, 13th Gen Intel(R) Core(TM) i9-13900H CPU.  
Training took approximately four hours and the ABC analysis took approximately 8 hours wall time.  
We generated 20,000 LCM simulations for training the SCNN and another 992,600 LCM simulations for the ABC analysis.  

\begin{table}[p]
    \centering
    \begin{tabular}{l c c}
         {\bfseries Layer type}           & {\bfseries Output shape}    & {\bfseries Parameters} \\
         \hline
         \texttt{Input}                   & $(N_b, \Npix, 1)$           & 0 \\
         \texttt{ChebyshevConv (K=32, Fout=40)}  & $(N_b, \Npix, 40)$    & 840 \\
         \texttt{MaxPool (p=2)}           & $(N_b, 12288, 40)$          & 0 \\
         \texttt{ChebyshevConv (K=10, Fout=20)}  & $(N_b, 12288, 20)$    & 8020 \\
         \texttt{MaxPool (p=1)}           & $(N_b, 768, 20)$            & 0 \\
         \texttt{ChebyshevConv (K=5, Fout=10)}   & $(N_b, 768, 10)$      & 1010 \\
         \texttt{GlobalAvgPool}           & $(N_b, 10)$                 & 0 \\
         \texttt{Flatten}                 & $(N_b, 10)$                 & 0 \\
         \texttt{Dense}                   & $(N_b, 1)$                  & 11 \\
         \hline
    \end{tabular}
    \caption{\label{tab:log-zeta-arch}
    Neural network architecture used for the $Z = \log_{10}(\zeta_0)$ estimator. The batch size is $N_b=32$ and the HEALPix resolution parameter is $\Nside=128$, so the number of pixels is $\Npix = 196,608$. All layers use ReLU activations and batch normalization is disabled (\texttt{use\_bn=False}).
    }
\end{table}

\begin{table}[p]
    \centering
    \begin{tabular}{l c c}
         {\bfseries Layer type}           & {\bfseries Output shape}    & {\bfseries Parameters} \\
         \hline
         \texttt{Input}                   & $(N_b, \Npix, 1)$           & 0 \\
         \texttt{ChebyshevConv (K=5, Fout=8)}   & $(N_b, \Npix, 8)$      & 48 \\
         \texttt{MaxPool (p=2)}           & $(N_b, 12288, 8)$           & 0 \\
         \texttt{ChebyshevConv (K=5, Fout=16)} & $(N_b, 12288, 16)$     & 656 \\
         \texttt{ChebyshevConv (K=5, Fout=16)} & $(N_b, 12288, 16)$     & 1296 \\
         \texttt{MaxPool (p=1)}           & $(N_b, 3072, 16)$           & 0 \\
         \texttt{ChebyshevConv (K=5, Fout=32)} & $(N_b, 3072, 32)$      & 2592 \\
         \texttt{ChebyshevConv (K=5, Fout=32)} & $(N_b, 3072, 32)$      & 5152 \\
         \texttt{MaxPool (p=1)}           & $(N_b, 768, 32)$            & 0 \\
         \texttt{ChebyshevConv (K=5, Fout=64)} & $(N_b, 768, 64)$       & 10304 \\
         \texttt{GlobalAvgPool}           & $(N_b, 64)$                 & 0 \\
         \texttt{Flatten}                 & $(N_b, 64)$                 & 0 \\
         \texttt{Dense}                   & $(N_b, 1)$                  & 65 \\
         \hline
    \end{tabular}
    \caption{\label{tab:log-A2xi0-arch}
    Neural network architecture used for the $A = \log_{10}(\mathcal{A}^2 \xi_0)$ estimator. The batch size is $N_b=32$ and the HEALPix resolution parameter is $\Nside=128$, so the number of pixels is $\Npix = 196,608$. All layers use ReLU activations and batch normalization is disabled (\texttt{use\_bn=False}).
    }
\end{table}

\begin{table}[p]
    \centering
    \begin{tabular}{l c c}
         {\bfseries Layer type}           & {\bfseries Output shape}    & {\bfseries Parameters} \\
         \hline
         \texttt{Input}                   & $(N_b, \Npix, 1)$           & 0 \\
         \texttt{ChebyshevConv (K=5, Fout=16)}  & $(N_b, \Npix, 16)$     & 96 \\
         \texttt{ChebyshevConv (K=5, Fout=32)}  & $(N_b, \Npix, 32)$     & 2592 \\
         \texttt{MaxPool (p=2)}           & $(N_b, 12288, 32)$          & 0 \\
         \texttt{ChebyshevConv (K=5, Fout=64)}  & $(N_b, 12288, 64)$     & 10432 \\
         \texttt{ChebyshevConv (K=5, Fout=128)} & $(N_b, 12288, 128)$   & 41344 \\
         \texttt{MaxPool (p=1)}           & $(N_b, 3072, 128)$          & 0 \\
         \texttt{ChebyshevConv (K=5, Fout=256)} & $(N_b, 3072, 256)$    & 164608 \\
         \texttt{ChebyshevConv (K=5, Fout=512)} & $(N_b, 3072, 512)$    & 656896 \\
         \texttt{MaxPool (p=1)}           & $(N_b, 768, 512)$           & 0 \\
         \texttt{ChebyshevConv (K=5, Fout=512)} & $(N_b, 768, 512)$     & 1312256 \\
         \texttt{ChebyshevConv (K=5, Fout=512)} & $(N_b, 768, 512)$     & 1312256 \\
         \texttt{GlobalAvgPool}           & $(N_b, 512)$                & 0 \\
         \texttt{Flatten}                 & $(N_b, 512)$                & 0 \\
         \texttt{Dense}                   & $(N_b, 128)$                & 65664 \\
         \texttt{Dense}                   & $(N_b, 1)$                  & 129 \\
         \hline
    \end{tabular}
    \caption{\label{tab:log-xi0_A-arch}
    Neural network architecture used for the $X = \log_{10}(\xi_0^2 / \mathcal{A})$ estimator. The batch size is $N_b=8$ and the HEALPix resolution parameter is $\Nside=128$, so the number of pixels is $\Npix = 196,608$. All layers use ReLU activations and batch normalization is disabled (\texttt{use\_bn=False}).
    }
\end{table}

\section{Additional ABC posteriors}
\label{app:additional}

\begin{figure}
    \centering
    \includegraphics[width=0.8\linewidth]{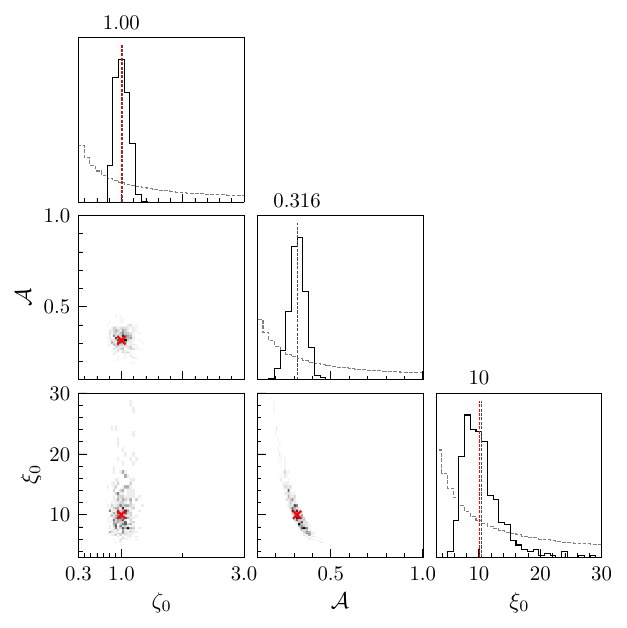}
    \caption{\label{fig:posterior_xi0A}
    Posterior plotted in terms of $\zeta_0$, $\mathcal{A}$, and $\xi_0$ for $\hat{\theta}_\mathrm{target} = (0, 0, 2.5)$ corresponding to $\hat{\zeta}_{0,\mathrm{targ}} = 1.0$, $\hat{\mathcal{A}}_{\mathrm{targ}} = 0.316$, $\hat{\xi}_{0,\mathrm{targ}} = 10$, which is the same as \fref{fig:posterior}. The samples used to produce this plot are exactly the same as the samples used to generate \fref{fig:posterior}. 
    }
\end{figure}

\begin{figure}
    \centering
    \includegraphics[width=0.8\linewidth]{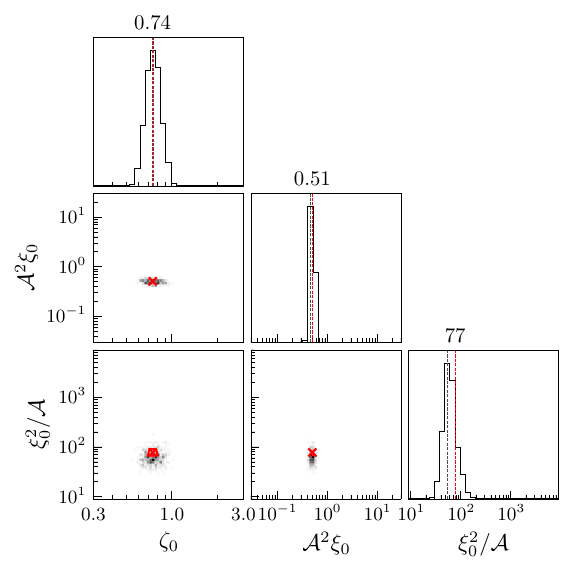}
    \caption{\label{fig:posterior_new_point}
    Similar to~\fref{fig:posterior} but with $\hat{\theta}_\mathrm{target} = (\hat{Z}$, $\hat{A}$, $\hat{X}) = (-0.13,\,-0.29,\,1.89)$.
    }
\end{figure}

In this appendix we provide two additional figures that show the posterior probability distributions that we obtain from ABC sampling.  
In the main body of the paper, the result of ABC sampling is presented in \fref{fig:posterior}.  
For that calculation we took the target point to be $\hat{\theta}_\mathrm{target} = (\hat{Z}$, $\hat{A}$, $\hat{X}) = (0,\,0,\,2.5)$ and we presented the posterior over the parameters $Z = \log_{10}(\zeta_0)$, $A = \log_{10}(\ampl)$, and $X = \log_{10}(\orthoAmpl)$.  
First, in \fref{fig:posterior_xi0A} we show the posterior in terms of $\log_{10} \zeta_0$, $\log_{10} \Acal$, and $\log_{10} \xi_0$.  
When the posterior is expressed in terms of $\Acal$ and $\xi_0$, the degeneracy direction is evident from the two-dimensional posterior.  
Second, in \fref{fig:posterior_new_point}, we provide posteriors obtained using ABC sampling with a different target point: $\hat{\theta}_\mathrm{target} = (\hat{Z}$, $\hat{A}$, $\hat{X}) = (-0.13,\,-0.29,\,1.89)$.  
As in \fref{fig:posterior} the diagonal subplots show the 1D marginal posteriors, gray vertical lines indicate the parameter with the highest posterior density, and red vertical lines indicate $\hat{\theta}_\mathrm{target}$.  
In the lower-left subplots, we show a histogram of 2D marginal posteriors with red crosses depicting the projection of $\hat{\theta}_\mathrm{target}$ in each plane.

\newpage
\bibliographystyle{JHEP}
\bibliography{references}

\end{document}